\documentstyle[11pt,newpasp,twoside,epsf]{article}
\def\edcomment#1{\iffalse\marginpar{\raggedright\sl#1\/}\else\relax\fi}
\reversemarginpar

\begin{document}
\title{Mapping Tidal Streams around Galactic Globular Clusters}
 \author{D. Mart\'\i nez-Delgado,$^{1,3}$ D. I. Dinescu,$^{2}$
  R. Zinn,$^{2}$ A. Tutsoff,${^3}$ P. C\^ot\'e,$^{4}$ and A. Boyarchuck${^3}$}
\affil{$^{1}$Max-Planck-Institut fur Astronomie, Heidelberg, Germany}
\affil{$^{2}$Department of Astronomy, Yale University, USA}
\affil{$^{3}$Instituto de Astrof\'\i sica de Canarias, La Laguna, Spain}
\affil{$^{4}$Department of Physics and Astronomy, Rutgers University, USA}
\setcounter{page}{255}
\begin{abstract}
The surroundings of globular clusters in the outer halo are one of the 
best places to look for the {\it ghost} of an ancient  dwarf galaxy. 
We  present the first preliminary results from a long term project 
designed to search for tidal debris around remote halo globular clusters, 
debris that may be the vestiges of the larger systems in which these 
clusters formed.

\end{abstract}

\section{Introduction}

Cold dark matter cosmology predicts that the halos of galaxies 
similar to the Milky Way form through the tidal disruption 
and accretion of numerous low-mass fragments---objects that 
may resemble the dwarf galaxies observed at the present time. 
The Milky Way offers us a unique laboratory 
to test this cosmological scenario by searching for
the fossil records of these merging events, which may take
the form of long  stellar streams or large scale stellar
substructure in the outer halo of our Galaxy
(see Bullock \& Johnston, this volume, p.\ 80). 

The surroundings of globular clusters in the outer halo are one of the best 
places to look for the remnants of ancient dwarf satellites. The classical Searle 
\& Zinn (1978)  scenario of the formation of the Milky  Way, in which 
the halo globular clusters formed in larger dwarf galaxies, can nowadays 
be considered the local manifestation of this hierarchical galaxy formation. 
It is now well known that some Galactic globular clusters are actually 
associated with the tidal stellar stream resulting from the  accretion 
of dwarf satellites, thereby confirming that this process plays an important 
role in the building of the Milky Way globular cluster system. Furthermore, 
several recent studies have suggested that the two most luminous Galactic 
globular clusters (M54, associated with Sagittarius, Sarajedini \& Layden 
1995,  and Omega Centauri, Lee et al. 1999) might be the nuclei of dwarf 
galaxies.  Extending previous suggestions, van den Bergh (2000) has argued 
that the  ``young'' globulars located in the outer halo might be the nuclei 
of extinct dwarf galaxies and suggests that a search for evidence of this 
may prove worthwhile.
If these clusters formed within larger stellar systems, we would expect to 
find them surrounded by a stellar population that should be revealed by 
deep photometry over a wide field. In this paper, we  present the first 
(most interesting) preliminary results from a long term project designed 
to search for tidal
debris around remote halo globular clusters, debris that may be the vestiges 
of the larger systems in which these clusters formed.

\begin{figure}  
\vspace{8cm}
\caption{Color--magnitude diagrams of confirmed tidal debris around globular
clusters: the remnants of the Sgr tidal stream around Pal 12 
(Mart\'\i nez-Delgado et al.\ 2002) and the prominent tidal tail 
of Pal 5 (Koch et al.\ 2004). The Sgr tidal stream population is 
also observed in the field around Pal 5 as a bump at $(B-R)\sim 
1,0$, $V\sim23$. {\itshape Upper panels:\/} The center of the 
clusters. {\itshape Lower panels:\/} The covered extratidal field.}
\end{figure}

\section{The Mapping of Tidal Streams with Color-magnitude diagrams}

Our main technique for mapping tidal streams in the Galactic halo is based
on the analysis of specific regions of the color--magnitude diagram (CMD)
 where the presence of the stellar population associated with the dwarf 
 satellite is expected as a density enhancement above the foreground 
 Galactic population, assuming the distance moduli of the parent galaxy 
 is the same that of the globular clusters.
Our previous experience in nearby systems (see Fig.\ 1) indicates that the 
best way to detect these structures through resolved stars is by obtaining
 observations deep enough to reach the  main sequence (MS)
turn-off. This prominent feature  is blue enough to avoid the contamination
 by Milky Way dwarf disk stars and is sometimes the only visible signature 
 of  the tidal debris of a dwarf galaxy in the CMD. Figure 1 shows some
examples of CMD morphologies of confirmed tidal debris around globular
 clusters (see also Mart\'\i nez-Delgado et al.\ 2004 for some model CMDs 
 of tidal streams).

All the data presented in this paper were obtained with the
Wide Field Camera at the 2.5~m Isaac Newton Telescope at 
Roque de los Muchachos Observatory (La Palma, Spain). Photometry was
obtained with the DAOPHOT II/ALLSTAR package. Transformation of our
photometry data to the standard system is still preliminary.

\section{The Survey}

\subsection {Looking for Ghostly Dwarf Galaxies around Globular Clusters}

 Our sample of selected globular clusters are members of the ``younger 
 halo'', which Zinn (1993)
proposed to be the accreted component after the collapse of the central
 regions of the
Galaxy. These dwarf remnants may be still undiscovered because of their
 very faint surface brightness ($\Sigma \sim$ 31 mag/arcsec$^{2}$) and scale 
 size (may be larger than $\sim$4 kpc because 
of its possible tidally disrupted state) and because these clusters have 
been observed only with small fields of view centered on the cluster. 

In addition, we also have included those globular clusters reported as 
possible members of known tidal streams in the Galactic halo.
It is now confirmed  that  Sgr and its tidal stream contains at least 
five globular clusters. The first target of this project, Pal 12, was 
confirmed to be a  {bona-fide} example of a  supposedly isolated 
globular cluster belonging to a tidal stream of this external galaxy 
(Dinescu et al.\ 2000; Mart\'\i nez-Delgado et al.\ 2002, see Fig.~1).
 A list of additional Sgr cluster candidates has been proposed by 
 Bellazzini et al.\ (this volume, p.\ 220). More recently,
Frinchaboy et al.\ (2004) have suggested that some nearby, 
low galactic latitude clusters are member of the Monoceros tidal 
stream (Yanny et al.\ 2003). Our main objective is to confirm 
whether  these clusters were associated in the past with these two 
dwarfs,  an important step towards understanding  the role of these merger
events in the building of the Galactic globular cluster
system. 
In addition, The CMD analysis of regions of these known tidal streams  
in the vicinity of these clusters also provides information on 
the distance, surface brightness, and stellar populations  of their
stellar debris (see Sects 4.1 and 4.2), properties that may be 
compared with $N$-body simulations of the tidal disruption of each 
particular satellite ( Mart\'\i nez-Delgado et al.\ 2004; 
Pe\~narrubia et al.\ 2004, in preparation).

\subsection {Tracing the Tidal Disruption of Globular Clusters}

Our deep observations  can reveal if some globular clusters have  
developed {\it tidal tails} of stars detached from the clusters 
by the tidal field of the Galaxy (see Grillmair et al.\ 1995; 
Odenkirchen et al.\ 2001). Consequently, this project also provides 
insights on the tidal disruption
of the clusters  even if they are not surrounded by a stellar 
population from a hypothetical host dwarf galaxy (see Sect.\ 4.3). 
Our deep photometry of tidal debris has already showed for first 
time evidence on the mass segregation in the tidal tail of a 
globular cluster ( Pal 5: Koch et al.\ 2004) or a possible clumpy 
distribution in the main body of Pal 5 (Mart\'\i nez-Delgado et 
al.\ 2004, in preparation).

\section{Results}

\subsection{Two Tidal Streams in the  Field around the Globular 
Cluster\\ NGC~4147/NGC~5024}

\begin{figure}[!t]
\vspace{15cm}
\caption{Color--magnitude diagrams of the central region of the Galactic 
globular clusters NGC 4147, NGC 7006, Pal 13 and Pal 14 ({\itshape left\/}) and 
their extratidal fields ({\itshape right\/}). The prominent blue clump observed 
in the majority of these diagrams at $(B-R)\sim 0.5$, $V\sim 24$ is due 
to the contamination of background galaxies (see
Mart\'\i nez-Delgado et al.\ 2004).}
\end{figure}

 Figure 2  shows the CMD  for 35$\arcmin \times$ 35$\arcmin$ field centered 
 on the vicinity of the globular cluster NGC 4147 ({\itshape left\/}) and
for an extratidal field ({\itshape right\/}), situated at 3 $^{\circ}$ from its 
center 
(and $\sim$ 2$^{\circ}$ from the globular cluster NGC 5024). This outer 
field shows two possible MS turn-off features, which could be related with
tidal debris of different systems or different wraps of a same stream. 
The MS feature with the turn-off at $B-R\sim$ 0.9 
and $V\sim 21.5$ and extending down to $V\sim 24$ corresponds to the 
Sgr northern tidal stream. The estimated heliocentric distance to this 
part of the stream from the MS turn-off magnitude is $\sim$40 kpc.

 A possible second MS turn-off is observed at $(B-R)\sim 1$ and $V\sim 
 20.5$, overlapping with the expected position of the cluster's MS 
 population. This could be related to a very low surface brightness, 
 additional wrap of the Sgr tidal stream situated at  a similar 
 distance to that of the cluster ($d\sim$ 20 kpc; see below). Another 
 explanation is that this brighter turn-off
 is the signature of the tidal tail of some of the globular clusters 
 in the vicinity (NGC 4147 or NGC 5024). However, our field does not 
 seem to lie along the paths of either cluster's orbits estimated from 
 their proper motions.
Finally, it could be also related to a new, very diffuse unknown 
tidal stream  detected towards the North Galactic Pole (see Newberg 
et al.\ 2002; Zinn et al.,
this volume, p.\ 92). Kinematic information of these MS star candidates is 
necessary to decide on the identification of this possible tidal debris.

\begin{table}[tbt]
\begin{center}
{\small
\begin{tabular}{rrrr}
\multicolumn{4}{c}{\small Table 1. Integrals of Motion} \\ \\
\hline
\noalign{\smallskip}
\multicolumn{1}{c}{Object} & 
\multicolumn{1}{c}{$E_{\rm orb}$}&
\multicolumn{1}{c}{$L_{z}$}&
\multicolumn{1}{c}{$L$} \\
& \multicolumn{1}{c}{($10^{4}$ km$^{2}$ s$^{-2}$)} &
\multicolumn{1}{c}{(kpc km s$^{-1}$)} &
\multicolumn{1}{c}{(kpc km s$^{-1}$)} \\
\noalign{\smallskip}
\hline
\noalign{\smallskip}
\multicolumn{1}{l}{Sgr$^1$} & \multicolumn{1}{r}{$-$0.8(1.1)} & \multicolumn{1}{r}{1074(396)} & \multicolumn{1}{c}{5365} \\
\multicolumn{1}{l}{Pal 12} & \multicolumn{1}{r}{$-$3.4(0.6)} & \multicolumn{1}{r}{1752(193)} & \multicolumn{1}{c}{3943} \\
\multicolumn{1}{l}{4147} & \multicolumn{1}{r}{$-$3.6(0.5)} & \multicolumn{1}{r}{$-$1187(322)} & \multicolumn{1}{c}{2733} \\
\multicolumn{1}{l}{5024} & \multicolumn{1}{r}{$-$1.9(1.5)} & \multicolumn{1}{r}{1263(290)} & \multicolumn{1}{c}{4809} \\
\noalign{\smallskip}
\hline 
\noalign{\smallskip}
\multicolumn{4}{l}{\small $^1$Based on a new determination from the SPM 3} \\
\multicolumn{4}{l}{\small catalog (Girard et al.\ 2004), and 2MASS).} \\
\end{tabular}
}
\end{center}
\end{table}

Another important question is whether NGC 4147 and/or NGC5024 could be
 formerly associated with the Sgr dwarf galaxy. Bellazzini et al.\ (2003) 
 argued for the association of
NGC 4147 with the Sgr stream from its radial velocity and from the 
detection of M giant  Sgr stars around this cluster. However, our 
deep photometry shows that the heliocentric distance of the Sgr 
northern stream  in the field is $\sim$20 kpc larger than the distance 
to NGC 4147 ($d=19.3$ kpc) or NGC~5024 ($d= 18.3$ kpc). 
Models show that in this region of the sky one can expect to see
two Sgr streams; one at $\sim$40 kpc which is a leading stream 
that eventually turns around and falls toward the solar neighborhood, 
and the second, more diffuse, one at
$\sim$20 kpc which is an old trailing stream.
Some of this is indeed seen in the field between NGC 4147 and NGC 5024
where two turn-offs are apparent: the more prominent one at $\sim$40 kpc
and the less populated one at $\sim$20 kpc. Thus even if a Sgr
stream at $\sim$20 kpc is present, it seems unlikely that such a diffuse 
stream
can contain two globular clusters that once belonged to the same system, 
especially given the two very different values of the radial velocities 
of the 
clusters.
Alternatively, Sgr debris in this part of the sky may be from the leading
stream turning around and falling toward the solar neighborhood, and 
therefore 
the clusters may belong to two different streams from Sgr.
Since the Sgr stream configuration becomes very complex in this region it
is hard to reach a definite conclusion regarding NGC 4147 and 5024.
We turn now to the proper motion information for these clusters.
In Table 1 we show the intergrals of motion for Sgr, for 
NGC 4147 and NGC 5024, and for Pal 12, the cluster now widely accepted to
have been part of Sgr. Orbits were integrated in the 
Johnson, Spergel, \& Hernquist (1995) potential, and proper motions
were taken from the recent literature\footnote{Sgr's proper motion is a new 
determination from the SPM 3 catalog 
(Girard et al.\ 2004) and 2MASS;this determination agrees with that
of Ibata et al.\ (1997), but has smaller uncertainties.}
 (Dinescu et al.\ 1999, 2000;
 Wang et al.\ 2000). Because of their large differences in $L_z$, the association
 of NGC~4147 with Sgr is unlikely. The agreement between NGC~5024 and Sgr
 is much better and is consistent with a dynamical association.

Our data also provide new insights into the stellar population of the Sgr
 northern stream observed in this field. The stellar population of both 
 clusters can be fitted very well with an old, metal-poor  stellar 
 theoretical isochrone ([Fe/H] $=-2.0$). This same isochrone fits the 
 population of the Sgr northern stream at the larger distance 
 ($d\sim 40$ kpc),
suggesting that this part of the stream could be very metal-poor, as also 
reported by Vivas, Zinn, \& Gallart (this volume, p.\ 108) 
for Sgr RR Lyrae stars. These 
results would indicate that the mean metallicity of the Sgr tidal 
stream in its apocenter region is more metal poor than those observed 
in its center region, as predicted by some theoretical models 
(see Mart\'\i nez-Delgado et al.\ 2004). However,  metallicity 
estimates from pure MS fitting of the tidal debris are fairly 
ambiguous, and more data (i.e., spectroscopic
 metal abundances of Sgr stream stars) are necessary 
to confirm this possible scenario.

\subsection{A Possible Older Wrap of the Monoceros Tidal Stream around 
NGC~7006}

NGC 7006 ($l=63.8^\circ, b=-19.4^\circ$)---an archetypical ``second parameter'' 
cluster---which resides in the outskips of the Milky Way, is
of considerable interest for formation scenarios of the assembling of the 
Galactic halo. The extratidal field of this cluster (Fig.\ 2) shows an MS 
feature at $(B-R)\sim 1.0$ and $V \sim 20.7$--23 that is not predicted by 
star count models of the Galactic foreground population computed from the 
TRILEGAL model of the Milky Way (Girard et al.\ 2004, in preparation) for 
these galactic coordinates.  Although it could be associated with the 
stellar component of some unknown Galactic perturbation, this population 
is more likely associated with a piece of a possible old, metal-poor stellar 
stream at a heliocentric distance of $\sim$25 kpc. The smaller distance 
of this possible tidal debris than that of
NGC 7006 ($d= 41.5$ kpc) suggests that the feature is not associated with this 
globular cluster. The redder color of its MS turn-off also suggests that 
its stellar population could be more metal-rich than that of the cluster 
([Fe/H] $\sim -1.6$).
The best candidate is the Tri/And stream  discovered in a large sky area 
at 17--20 kpc (Majewski et al.\ 2004), which
shows a similar MS turn-off feature in the CMDs. Our new detection at 
lower galactic longitude indicates that the distance to this tidal stream 
increases toward the Galactic center, in good
agreement with the predictions of theoretical models that shows this tidal 
debris as an older, more distant wrap of the Monoceros tidal stream 
(Pe\~narrubia et al.\ 2004, in preparation).

\subsection{The Tidal Disruption of Palomar 13}

Pal 13 is a low luminosity outer halo cluster in  an advanced state of
 total tidal dissolution by the
Milky Way (Siegel et al.\ 2001). The actual cluster size is controversial, 
so we derived its structural
parameters using our deep wide field data as a first step to the search 
for tidal debris in its vicinity.
Figure 3 shows the density profile of the cluster obtained from our INT data, 
in comparison with
those obtained by C\^ot\'e et al.\ (2002). This profile shows a pronounced 
``inflection'', as predicted by $N$-body simulations of disrupting satellites or
observed in the Sgr main body (Majewski et al.\ 2003), providing strong
observational evidence about its tidal destruction. Our best-fit King--Michie
 model (dashed line in Fig.\ 3) yields a tidal radius of $r_{\rm t}= 17\arcmin$, 
 confirming the suggestion that the cluster is considerably more extended
  than previously suspected (C\^ot\'e et al.\ 2002). This conclusion
is supported by the detection of at least one probable cluster member at
$\sim$15$\arcmin$ from the cluster center, found in a radial velocity 
survey
conducted with the ESI spectrograph on the 10~m Keck  Telescope.

The CMD of our control field (situated at $\sim$2.5$^{\circ}$) shows a
 possible MS feature with position and slope  coincident with those of 
 the cluster (Fig.~2). There are also two possible diffuse clumps of red 
 stars  at $((B-R),V )\sim (1.6, 18.5)$ and (1.4, 20) respectively that 
 overlap with the expected position of the subgiant branch of Pal 13. 
 These features could be related with the foreground halo population, 
 although our comparison with different CMD  models of the Milky Way 
 population (Bensacon, TRILEGAL) prevent us from reaching a definitive 
 conclusion. However, this  CMD morphology resembles those observed 
 in the tidal tails of the globular cluster Pal 5 (see Fig.\ 1), but with a 
 significantly lower surface brightness, suggesting they could be related 
 to the remmant of a substantially extended extratidal cluster population.
  The presence of such tidal extension would be consistent with the 
  elongated shape of the isodensity contours of Pal 13 in the direction
   close to the cluster's proper motion obtained from our data. 
   Unfortunately, the position of our control field is not aligned with 
   the cluster's proper motion, where the maximum stellar density from 
   the possible tidal tail emanating from the clusters would be expected.
A new wider survey around Pal 13 to confirm the presence of this possible 
relic of the cluster population is in progress. 

\begin{figure}[!ht] 
\plotfiddle{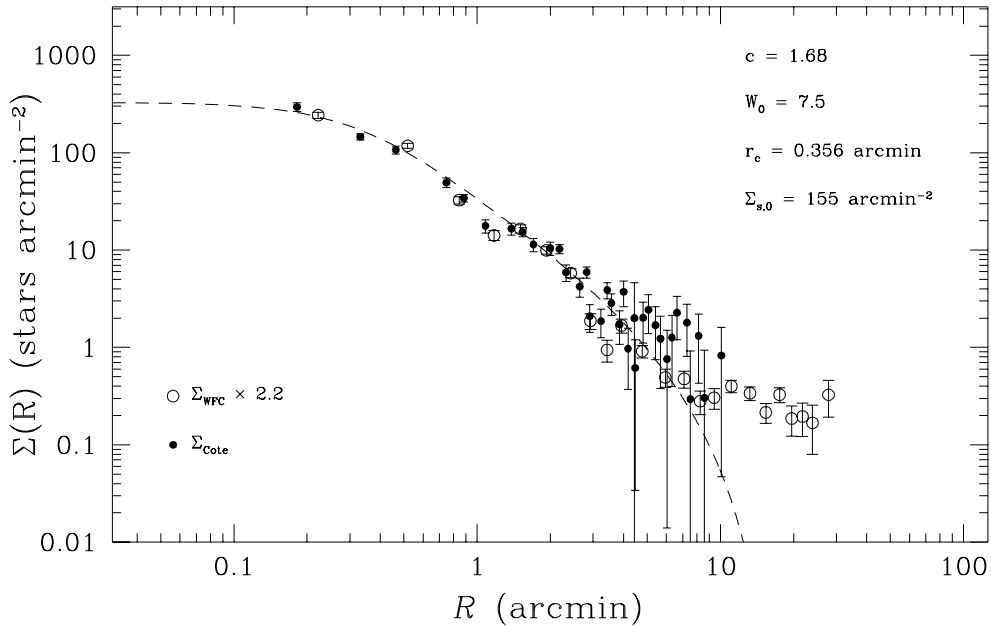}{8cm}{0}{65}{65}{-200}{-120}
\vspace{1cm}
\caption{Comparison of the surface density profile of Pal
13 given in C\^ot\'e et al.\ (2002) (filled circles) with our INT-WFC
data (open circles).}
\end{figure}

\subsection{An Extratidal Stellar Population around Palomar 14?}

It has been confirmed that Pal 14 is significantly younger than the average 
age of halo globular clusters (Sarajedini 1997) and is therefore one 
of the
members of the Zinn's ``younger halo'' clusters.  Its large galactocentric 
distance ($R> 70$ kpc) suggests that the cluster has probably completed very 
few orbits around the Galactic Center. 

Figure 2 ({\itshape right\/}) shows the CMD of the extratidal region (assuming a 
tidal
radius  for Pal 14 of $r_{\rm t}=5.2\arcmin$) for a total area of 0.26 
deg$^{2}$ around
the cluster. It still shows
a hook-like feature that is the vestige of the MS sequence of the cluster
at $(B-R)\sim 0.9$ and $V\sim 23.5$. The total extension of this
extratidal population is very uncertain because of the severe
background galaxy contamination in this region of the CMD, but it could 
extend out to $2\times r_{\rm t}$  from the center of the cluster. These 
extratidal
stars could be debris from the cluster or stars from a surrounding larger
system, but more observations are needed to reach a definitive conclusion 
concerning its origin.

\acknowledgements RZ was supported by NSF grant AST-0098428.


\end{document}